# High spin polarization in CoFeMnGe equiatomic quaternary Heusler alloy

Lakhan Bainsla,[1,4] K. G. Suresh,[1, #] A. K. Nigam,[2] M. Manivel Raja,[3] B.S.D.Ch.S. Varaprasad,[4] Y. K. Takahashi,[4] and K. Hono[4]

[1]Department of Physics, Indian Institute of Technology Bombay, Mumbai 400076, India
[2]Tata Institute of Fundamental Research, Homi Bhabha Road, Mumbai 400005, India
[3]Defence Metallurgical Research Laboratory, Hyderabad 500058, India
[4]Magnetic Materials unit, National Institute for Materials Science, Tsukuba 305-0047, Japan

**Abstract**

We report the structure, magnetic property and spin polarization of CoFeMnGe equiatomic quaternary Heusler alloy. The alloy was found to exist in the $L2_1$ structure with considerable amount of $DO_3$ disorder. Thermal analysis result indicated the Curie temperature is about 711K without any other phase transformation up to melting temperature. The magnetization value was close to that predicted by the Slater-Pauling curve. Current spin polarization of $P = 0.70 \pm 0.1$ was deduced using point contact Andreev reflection (PCAR) measurements. Half-metallic trend in the resistivity has also been observed in the temperature range of 5 K to 300 K. Considering the high spin polarization and Curie temperature, this material appears to be promising for spintronic applications.

Corresponding author (#: suresh@phy.iitb.ac.in )

Co-based Heusler alloys have attracted a lot of attention due to their theoretically predicted half-metallic nature and experimentally observed high spin polarization values with high Curie temperatures.[1-6] Half-metallic ferromagnets (FM) have applications as spin ploraized current sources for magnetic tunneling junctions,[7] current-perpendicular-to-plane giantmagnetoresistive (CPP-GMR) devices,[8-9] spin injectors to semiconductors,[10] and lateral spin valves.[11] Recently reported $Co_2Fe(Ga_{0.5}Ge_{0.5})$,[3] and $Co2(Fe_{0.4}Mn_{0.6})Si$[7] show rather high spin polarization at room temperature; however, their spin transport properties at room temperature is substantially degraded. Because of this, further exploration of half-metallic Heusler alloys is strongly desired.

In ternary Heusler alloys [*XYZ* (half Heusler) and *X₂YZ* (full Heusler)], *X* and *Y* are transition and *Z* is a main group element. Full Heusler alloys have the $L2_1$ structure (space group No. # 225) consisting of four interpenetrating sub-lattices. If each of the four sublattices is occupied by different atoms, a quaternary Heusler structure with different symmetry (space group no. # 216) is obtained, i.e., LiMgPdSn type.[12, 13] Depending on the occupation of the different lattice sites, three different type of structure is possible for the LiMgPdSn structure. To the best of our knowledge only a few quaternary Heusler alloys with 1:1:1:1 stoichiometry have been explored.[14,15]

Heusler alloys are known to exhibit tunable electronic and magnetic properties depending on their valance electron number. The usual method to conceive new quaternary Heusler alloys has been to combine two ternary Heusler alloys such as $Co_2FeGe$[16] and $Co_2MnGe$[17], which results in the quaternary CoFeMnGe (CFMG) alloy.[14] This alloy has been predicted to have a half-metallic nature with a high Curie temperature[16]. Recently some other quaternary Heusler alloys, CoFeMnX (X = Al, Si, Ga, Ge)[14,15] have been synthesized successfully. *Ab initio* calculations as well as experimental observations indicate that there is great possibility of achieving half-metallicity in these alloys. Klaer et al.[15] have predicted half-metallic behavior in the case of CFMG. For applications, such quaternary Heusler alloys with 1:1:1:1 stoichiometery have advantage over pseudo-ternary alloys such as $Co_2Fe(Ga_{0.5}Ge_{0.5})$, because the random distribution of Ga and Ge in $Co_2Fe(Ga_{0.5}Ge_{0.5})$ leads to an additional disorder scattering and thus the spin diffusion length becomes very short, only in the order of a few nm.[18] On the other hand, the devices based on equiatomic quaternary Heusler alloys are expected to have lower power dissipations. In the light of these, we have investigated the structural, magnetic, transport and current spin polarization properties of equiatomic CFMG quaternary Heusler alloy.

The polycrystalline bulk sample of an equiatomic CoFeMnGe (CFMG) alloy was prepared by arc melting of stoichiometric quantities of constituent elements in an inert atmosphere. The ingot was melted several times to increase the chemical homogeneity and the final weight loss was less than 1%. To increase the homogeneity, the as-cast alloy was annealed under vacuum for 14 days at 1073 K and then quenched in cold water. The crystal structure was investigated by x-ray diffraction with Cu K$_\alpha$ radiation at room temperature. $^{57}$Fe Mössbauer spectra at room temperature were recorded using a constant acceleration spectrometer with 25 mCi $^{57}$Co(Rh) radioactive source. The spectrometer was calibrated by using natural iron foil of 25 μm before measuring the sample. The obtained spectra were analyzed using PCMOS-II least-squares fitting program. Curie temperature and the structural transition temperature (if any) were probed by using the differential thermal analysis (DTA) with a heating rate of 20 K min$^{-1}$. Magnetization measurements at 3 K and 300 K were performed by using a vibrating sample magnetometer (VSM) attached to a physical property measurement system (PPMS, Quantum design). Current spin polarization measurements were done by using the point contact Andreev reflection (PCAR) technique.[19] Sharp Nb tips prepared by electrochemical polishing were used to make point contacts with the sample. Spin polarization of the conduction electrons was obtained by fitting the normalized conductance G(V)/G$_n$ curves to the modified Blonder-Tinkham-Klapwijk (BTK) model.[20] A multiple parameter least squares fitting was carried out to deduce spin polarization (*P*) using dimensionless interfacial scattering parameter (*Z*), superconducting band gap (*Δ*) and *P* as variables.

Fig. 1 shows the Rietveld refinement of powder XRD pattern recorded at room temperature. The lattice parameter of the alloy was found to be 5.75 Å, which is in agreement with the earlier reports on CFMG.[14] As one can see, superlattice reflections, (111) and (200), are very weak for this alloy. When *Z* element is from the same period of the periodic system as the transition metals, it is very difficult to find out correct structure unambiguously by using x-ray or neutron diffraction data. It is reported that (111) and (200) reflections were not observed in the XRD patterns of the Co$_2$FeZ (Z=Al, Si, Ga, Ge) samples, but the extended x-ray absorption fine structure technique showed the existence of L2$_1$ order[16]. It may be noted that the correct L2$_1$ structure is a necessary requirement for high spin polarization[21]. In order to make a rough estimate of chemical order present in the alloy, we calculated the $I_{200}/I_{220}$ and $I_{111}/I_{220}$ intensity ratios. The superlattice reflections L2$_1$ (111) and B2 (200) are proportional to the order parameters as $S^2$ and $S^2(1-2\alpha)^2$, where $S = ((I_{200}/I_{220})_{Exp}/(I_{200}/I_{220})_{Theory})^{1/2}$ and $S(1-2\alpha) = ((I_{100}/I_{220})_{Exp}/(I_{100}/I_{220})_{Theory})^{1/2}$.[22] In the case of A2 disorder, S = 0 and α = 0, for complete B2 order (without L2$_1$ order) S = 1 and α = 0.5. For the L2$_1$ order, S = 1 and α = 0 are expected. Therefore, the x-ray diffraction analysis gives an indication that the alloy has the L2$_1$ structure.

To further investigate the structural order, we performed $^{57}$Fe Mössbauer spectroscopic measurements at room temperature as shown in Fig. 2. The experimental spectrum has been fitted with two sextets having hyperfine magnetic field values of 285 and 104 kOe and relative intensities of 67 and 33% respectively. The quadrupole splitting and isomer shift values are very small. The data has been analyzed by taking the L2$_1$ ordering and type II LiMgPdSn structure, where Co, Mn, Fe and Ge occupy $X, X`, Y$ and $Z$ site respectively in quaternary Heusler alloy.[13, 15] It is reported from theoretical calculations that quaternary CoFeMnSi Heusler alloy can be half-metallic for type I and type II structure.[13] Two sextets namely S$_1$ and S$_2$ are found to be essential for obtaining good fitting. In an ideal L2$_1$ ordered structure, Fe atoms must occupy the $Y$ sites with cubic symmetry (O$_h$), which will result in a single sextet because there is only one crystallographic site for Fe. The presence of the second sextet could be attributed to the occupation of Fe atoms at $X$ or $Z$ sites, thereby indicating some amount of structural disorder. When Fe occupies $Z$ site, hyperfine magnetic field is expected not to change much as the number of magnetic near neighbours are similar for $Y$ and $Z$ sites whereas when it goes to $X$ site, a large decrease in hyperfine magnetic field is expected as this has the highest number of non-magnetic near neighbours. The experimentally observed values of hyperfine magnetic fields (285, 104 kOe) clearly indicate that the Fe also occupy $X, X'$ sites, resulting in DO$_3$ type disorder. Now, the sub-spectra S$_1$ and S$_2$ are ascribed to L2$_1$ and DO$_3$ phases respectively. The intensity of S$_1$ (67%) is found to be higher as compared to S$_2$ (33%), which means that the structure is reasonably ordered at room temperature. The values of the hyperfine field (H$_{hf}$) and isomer shift for spectra S$_1$ are comparable to those obtained for other Co-based Heusler alloys.[23, 24] quadrupole shift value is almost zero, which is in accordance with the cubic symmetry of the local Fe environment.

In order to investigate the structural stability with temperature, we performed differential thermal analysis (DTA) in the temperature range from 400 K to 1450 K as shown in Fig.3. In the DTA curve, there are only two minima, one is near $T_C$ and the other close to $T_m$ (melting temperature), which suggests there is no structural transition in the CFMG alloy up to the melting tempereature. Hence the alloy is ordered on solidification, i.e. it is an intermetallic compound, which is good from the point of view of applications. Curie temperature of ~750 K obtained from the DTA curve agrees very well with the report by Alijani et al.[14]

Isothermal magnetization measurements have been done at various temperatures upto a field of ±40 kOe and are shown in Fig.4. The M-H curves show the characteristics of a soft ferromagnet. Saturation magnetization of 4.2 $\mu_B$/f.u. and 3.9 $\mu_B$/f.u. are observed at 3 K and 300 K respectively, which are in good agreement with earlier observations.[14, 15] According to the Slater-Pauling rule[25], the value of saturation magnetization is 4 $\mu_B$/f.u. The experimental value is higher than the theoretical value, which may be due to the presence of the $D0_3$ structural disorder as observed in Mössbauer spectroscopy.

Spin polarization of the sample was measured by using the PCAR technique. In the measurement of spin polarization by PCAR, one measures the conductance curves across the ferromagnetic (FM)/superconductor (SC) point contact. Since spin polarization ($P_C$) measured by the PCAR is the transport spin polarization and hence it is much more important from the application point of view in spintronic devices such as CPP-GMR devices. Fig. 5(a)-5(c) shows the normalized conductance curves with typical bias dependence. In Fig. 5(a)-5(c) open circles denote the experimental data and the solid lines are fit to the data by modified BTK model.[20] As one can see, the shape of the curves change near the superconducting band gap ($\Delta$) with change in interfacial scattering parameter ($Z$). We fitted the measured data using modified BTK model by keeping $P$, $\Delta$ and $Z$ as variables. Since there is no proximity effect observed in the PCAR spectra, we assumed $\Delta = \Delta_1 = \Delta_2$. The values of $P$, $\Delta$ and $Z$ with the best fitting (least $\chi^2$ value) are shown in the figures. The conductance curves become more flat near $\Delta$ for low $Z$ values. This can be easily observed by closely looking at Fig. 5(a) [with $Z = 0.13$] and 5(c) [$Z = 0.29$]. $\Delta$ values obtained from the best fit is lower than the bulk superconducting band gap of Nb (1.5 meV), which is attributed to the multiple contacts which can give rise to the suppression of the band gap as reported for Nb/Cu by Clowes et al.[26] The value of current spin polarization is estimated by obtaining the conductance curves for $Z = 0$, but in our case the lowest value of $Z$ was 0.13 and hence the current spin polarization was estimated by linear fitting of the experimental data and extrapolating it to $Z = 0$. Fig. 5(d) shows the $P$ vs. $Z$ curve with extrapolation down to $Z = 0$.

The current spin polarization value of $0.70 \pm 0.01$ is deduced from the $P$ vs. $Z$ plot. The high value of spin polarization obtained for this bulk alloy is comparable to that obtained for bulk $Co_2Fe(Ga_{0.5}Ge_{0.5})$ ($P = 0.69 \pm 0.02$),[3] which has been proved to be one of the best FM Heusler alloy for CPP-GMR devices.[9, 11, 18, 27] This is the first report on the experimental measurement of the spin polarization in equiatomic quaternary Heusler alloys.

Electrical resistivity measurements have been done in 0 and 50 kOe in the temperature range of 5- 300 K as shown in Fig. 6. Resistivity value of 3.1 µΩm are found for CFMG which is relatively high as compared to that of $Co_2Fe(Ga_{0.5}Ge_{0.5})$ (0.6 µΩm) alloy,[18] which may be due to the presence of larger $DO_3$ disorder. Zero field resistivity has been normalized with respect to the value at 5 K and plotted in Fig. 6 (right hand side scale). The residual resistivity ratio [$\rho(T)/\rho_{5K}$] at 300 K is found to be low (1.02), which confirms the large disorder in the alloy. A minimum in the resistivity curve is found near 40 K, which is followed by an upturn at lower temperatures. Such a feature has been reported in many Heusler alloys[28] and is typically attributed to the disorder enhanced coherent scattering of conduction electrons.[29] Resistivity behavior has been analyzed in the temperature range of 5 - 300 K by considering different scattering mechanisms. Typically half-metallic Heusler alloys exhibit electron-electron ($T^2$ dependence)[30], electron-phonon (T dependence)[30] and double magnon scattering ($T^{9/2}$ at low temperatures and $T^{7/2}$ at high temperatures)[31, 32] dominated temperature regimes. While electron-magnon scattering is absent in the case of true half-metals, $T^{1/2}$ dependence reflects the contribution due to the disorder.[28] Fitting of the resistivity data has been carried out in two different temperature regions, i.e., between 5 and 85 K and between 85 and 300 K. In the low temperature region, the resistivity fits well with the relation, $\rho(T) = \rho_0 - AT^{1/2} + BT^2 + CT^{9/2}$, while in the high temperature region, the best fit is given by $\rho(T) = \rho_0 + DT + ET^{7/2}$. In the low temperature regime, $T^{1/2}$ term dominates over the other two contributions due to the presence of disorder. The linear dependence due to the electron-phonon contribution dominates in the high temperature regime with a weak negative contribution from the $T^{7/2}$ dependency. Thus the resistivity measurements show the signature of half-metallicity even at high temperatures. The values of various parameters obtained from the fits are given in table 2.

In conclusion, the quaternary equiatomic Heusler alloy CoFeMnGe was found to exist in the $L2_1$ structure with considerable amount of $DO_3$ disorder. No disordering transition was detected up to the melting temperature, indicating the alloy is a $L2_1$ intermetallic compound. Magnetization values roughly agree with those calculated by Slater-Pauling rule. The value of the current spin polarization was deduced to be $0.70 \pm 0.1$ by PCAR measurements at 4K, which is one of the highest values for ternary or quaternary Heusler alloys reported so far.[5,6,33] Electrical resistivity measurements shows the signature of half-metallicity for the material even at high temperatures. Therefore, this material appears to be promising to be used in spintronic devices such as GMR, which need high conductive P to achieve high MR ratios.

**Acknowledgement**

One of the authors, Lakhan Bainsla would like to thank UGC, New Delhi for granting Senior research fellowship (SRF). KGS thanks ISRO cell, IITB for funding the work.

**Figure Captions**

Fig.1. Rietveld refinement of the powder x-ray diffraction pattern of CoFeMnGe collected at the room temperature. The inset shows the expanded part of the low angle region.

Fig. 2. $^{57}$Fe Mössbauer spectra of CoFeMnGe collected at room temperature.

Fig. 3. Temperature variation of the DTA curve of CoFeMnGe in the temperature range of 400 K to 1450 K.

Fig. 4. Isothermal magnetization curves of CoFeMnGe at 3 K and 300 K. Solid line denotes the magnetization value obtained from Slater-Pauling rule.

Fig. 5. PCAR normalized conductance curves in CoFeMnGe measured at 4.2 K by using Nb as a superconducting tip. Solid lines are fit to the experimental data by modified BTK model and open circles are experimental data. (d) P vs. Z plots with extrapolation down to Z = 0.

FIG. 6. Electrical resistivity measurements under the field of 0 kOe and 50 kOe in the temperature range of 5 K to 300 K (left hand side scale). Residual resistivity ratio (RRR) vs. temperature curve right (hand side scale).

**Table.1**. Hyperfine field ($H_{hf}$), quadrupole splitting ($\Delta E_q$), isomer shift ($\delta$), line width and relative intensity of Mössbauer spectra collected at room temperature.

|  | Sub Spectrum | Mossbauer Parameters | | | | Phase Identified |
|---|---|---|---|---|---|---|
|  |  | $H_{hf}$ (kOe) | Quadrupole Shift ($\Delta E_q$) (mm/s) | Isomer Shift $\delta$ (mm/s) | Relative Intensity (%) |  |
| CFMG | $S_1$ | 285 | -0.011 | -0.003 | 67 | $L2_1$ (Y-site) |
|  | $S_2$ | 104 | -0.115 | 0.179 | 33 | $DO_3$ (X-site) |

**Table. 2.** Analysis of the zero field electrical resistivity in the temperature range of 5 K to 300 K.

| $\rho$ (5 K) | $\rho$ (300 K) | RRR | Fitting Parameters | | | | | | |
|---|---|---|---|---|---|---|---|---|---|
|  |  |  | Region I | | | | Region II | | |
| 3.09 µΩ.m | 3.16 µΩ.m | 1.02 | $\rho_0$ (µΩ.m) | A (µΩ.mK$^{-1/2}$) × 10$^{-3}$ | B (µΩ.m K$^{-2}$) × 10$^{-6}$ | C (µΩ.mK$^{-9/2}$) × 10$^{-11}$ | $\rho_0$ (µΩ.m) | D (µΩ.m K$^{-1}$) × 10$^{-4}$ | E (µΩ.mK$^{-7/2}$) × 10$^{-11}$ |
|  |  |  | 3.09 | 2.29 | 2.72 | 1.04 | 3.06 | 4.35 | -5.69 |

**Figures**

Fig. 1.

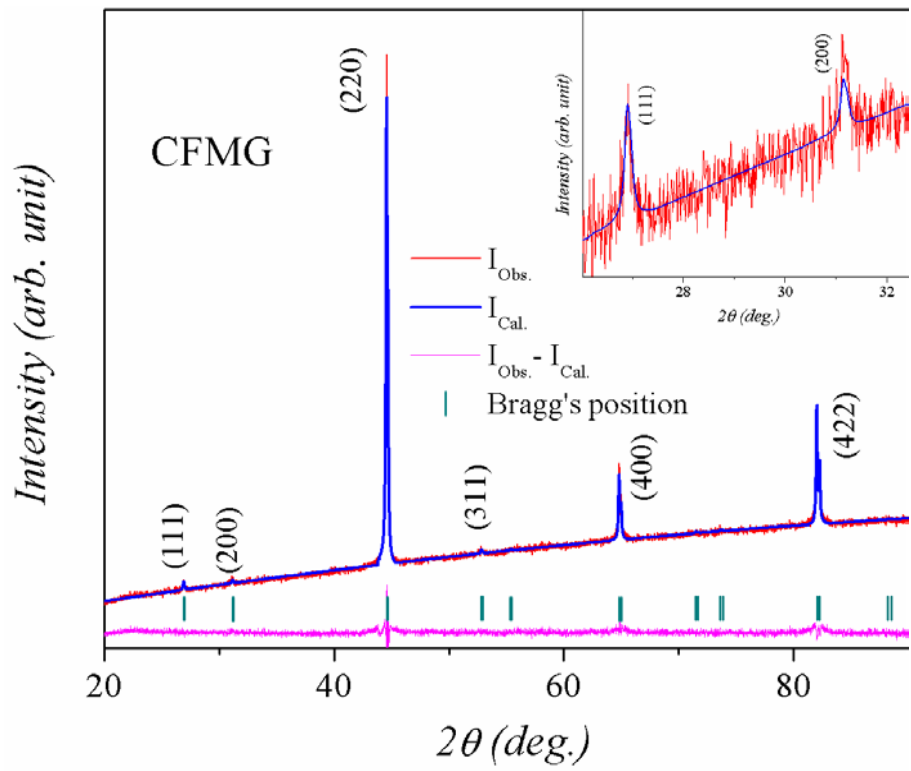

Fig. 2.

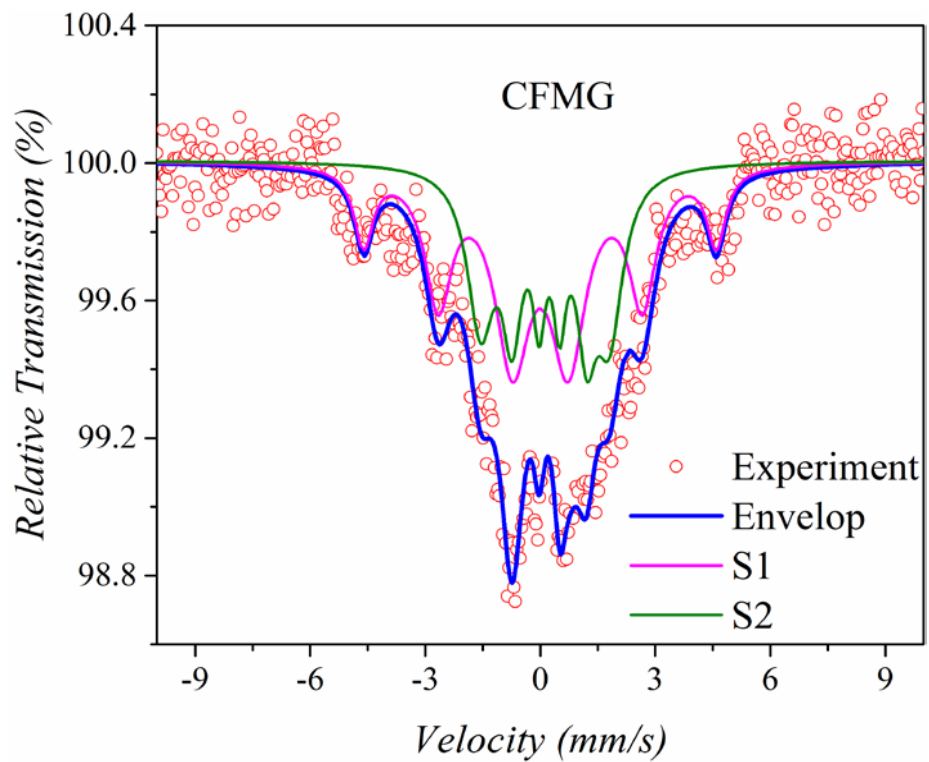

Fig. 3.

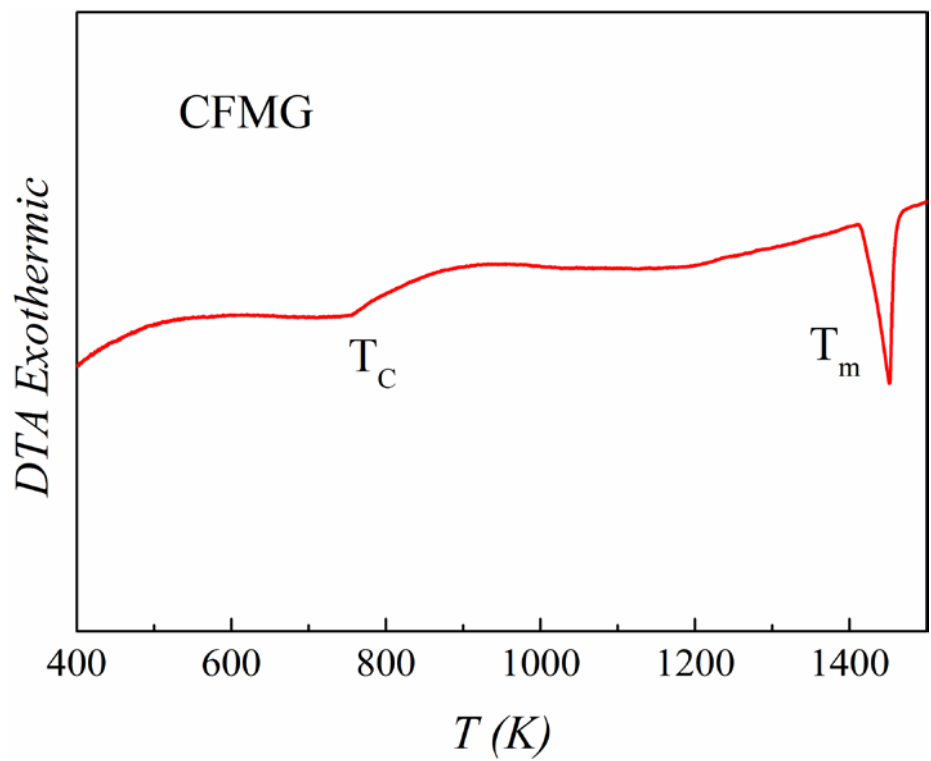

Fig. 4.

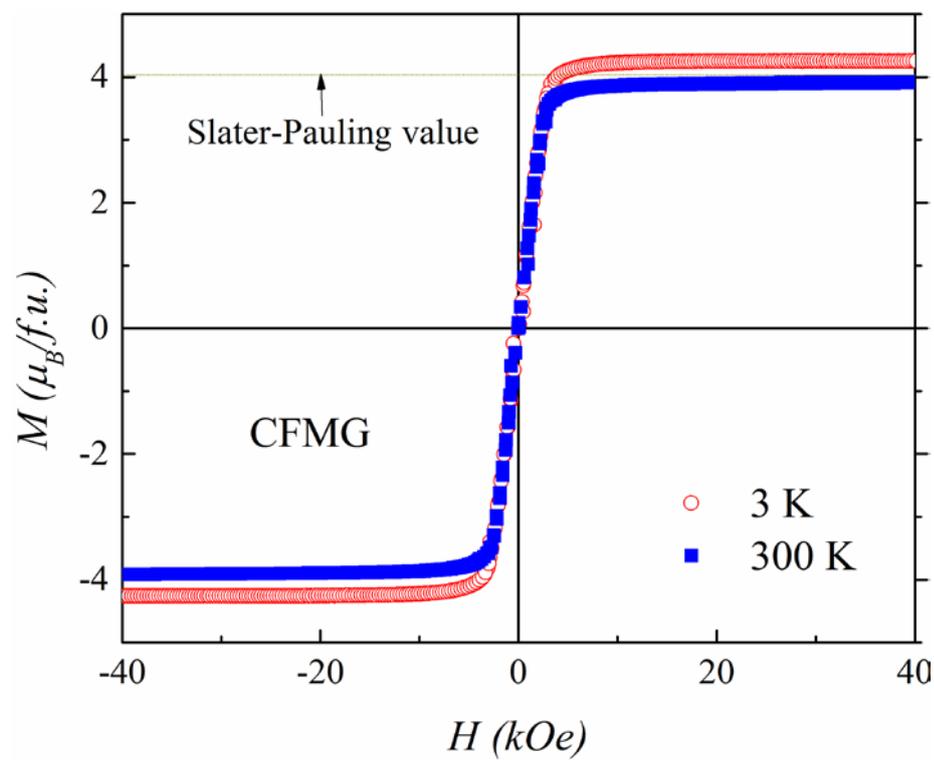

Fig. 5.

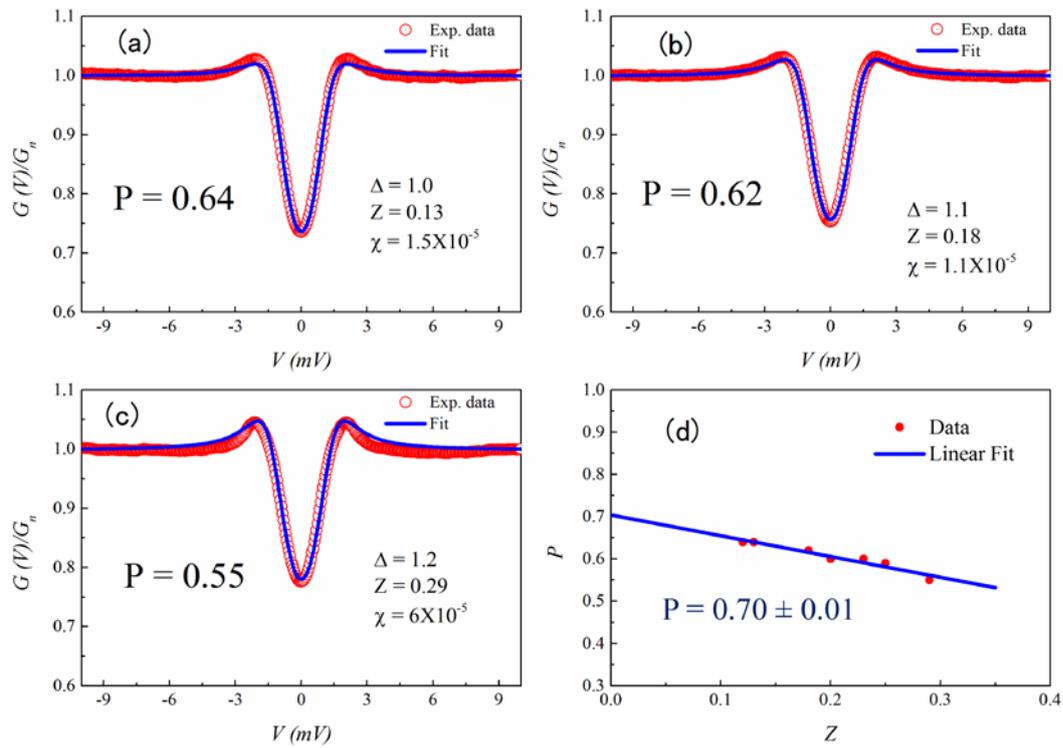

FIG.6.

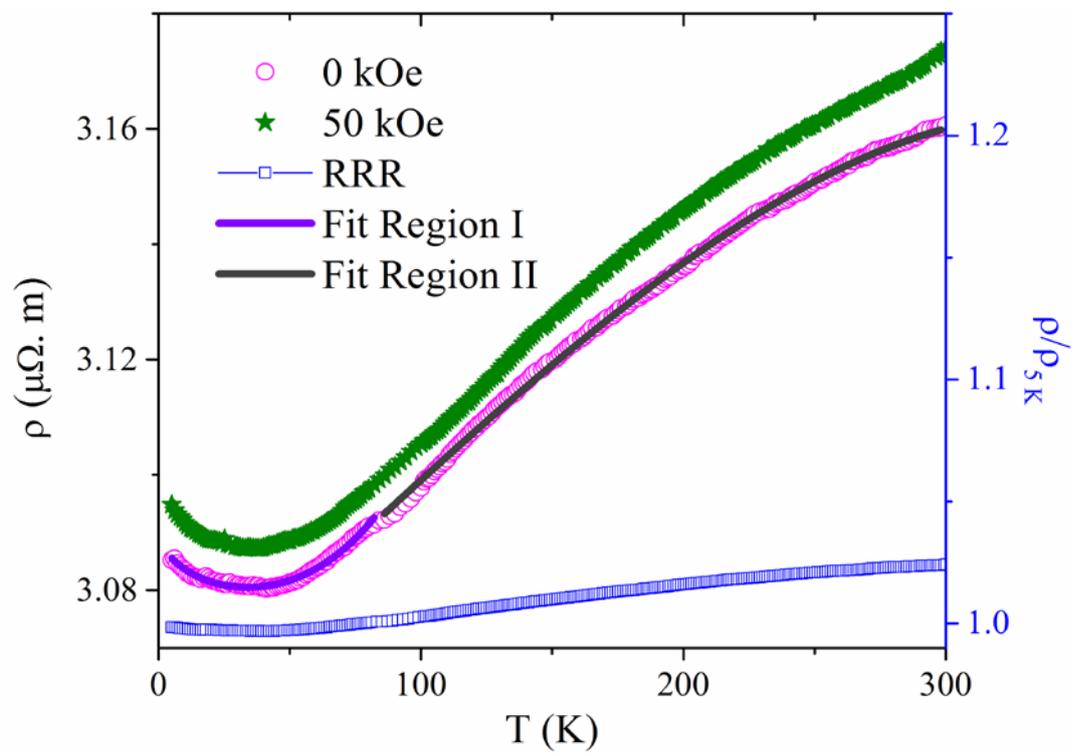